\documentclass[5p,twocolumn]{elsarticle}

\usepackage{algorithm}
\usepackage[noend]{algpseudocode}
\usepackage{amsmath}
\usepackage{amssymb}
\usepackage{color}
\usepackage{comment}
\usepackage{epstopdf}
\usepackage{graphicx}
\usepackage{hyperref}
\usepackage{soul}
\usepackage{tabularx}
\usepackage{tikz}
\usepackage{widetext}

\usetikzlibrary{shapes,arrows}
\tikzstyle{startstop} = [rectangle, rounded corners, minimum width=3cm, minimum height=1cm, text centered, draw=black]
\tikzstyle{process} = [rectangle, minimum width=3cm, minimum height=1cm, text centered, draw=black]
\tikzstyle{decision} = [diamond, minimum width=2cm, minimum height=1cm, text centered, draw=black]
\tikzstyle{arrow} = [thick,->,>=stealth]

\biboptions{sort&compress}

\journal{Computer Physics Communications}

\newcommand\pluseq{\mathrel{+}=}
\newcommand\plusplus{\mathrel{++}}
\newcommand\andeq{\mathrel{\&}=}
\newcommand\mathand{\mathrel{\&}}
\newcommand\mathmod{\mathrel{\%}}

\algnewcommand{\LineComment}[1]{\State\(\triangleright\) #1}
\algnewcommand\algorithmicinput{\textbf{Input:}}
\algnewcommand\Input{\item[\algorithmicinput]}
\algnewcommand\algorithmicoutput{\textbf{Output:}}
\algnewcommand\Output{\item[\algorithmicoutput]}

\makeatletter
\newcommand{\algmargin}{\the\ALG@thistlm}
\makeatother
\newlength{\ifwidth}
\newlength{\elseifwidth}
\algdef{SE}[parIF]{parIf}{EndparIf}[1]
  {\parbox[t]{\dimexpr\linewidth-\algmargin}{%
    \hangindent\ifwidth\strut\algorithmicif\ #1\ \algorithmicthen\strut}}{\algorithmicend\ \algorithmicif}
\algdef{C}[parIF]{parIF}{parElseIf}[1]
  {\parbox[t]{\dimexpr\linewidth-\algmargin}{%
    \hangindent\elseifwidth\strut\algorithmicelse\ \algorithmicif\ #1\ \algorithmicthen\strut}}
\algdef{Ce}[parELSE]{parIF}{parElse}{EndparIf}{\algorithmicelse}
\algnewcommand{\parState}[1]{\State%
  \parbox[t]{\dimexpr\linewidth-\algmargin}{\strut #1\strut}}
\algnewcommand{\parComment}[1]{\LineComment%
  \parbox[t]{\dimexpr\linewidth-\algmargin}{\strut #1\strut}}

\let\oldfootnote\footnote
\def\footnote{\ifhmode\unskip\fi\oldfootnote}

\newcommand{\blue}[1]{{\color{blue}{#1}}}
\renewcommand{\blue}[1]{{#1}}
\let\oldst\st
\renewcommand{\st}[1]{\blue{\oldst{#1}}}

\begin{document}

\begin{frontmatter}
\title{Causal Set Generator and Action Computer}
\author{William J.\ Cunningham}
\ead{w.cunningham@northeastern.edu}
\address{Department of Physics, Northeastern University, \\360 Huntington Ave. Boston, MA 02115, United States}
\author{Dmitri Krioukov}
\ead{dima@northeastern.edu}
\address{Departments of Physics, Mathematics, and Electrical \& Computer Engineering, Northeastern University, \\360 Huntington Ave. Boston, MA 02115, United States}

\begin{abstract}
The causal set approach to quantum gravity has gained traction over the past three decades, but numerical experiments involving causal sets have been limited to relatively small scales. The software suite presented here provides a new framework for the generation and study of causal sets. Its efficiency surpasses previous implementations by several orders of magnitude. We highlight several important features of the code, including the compact data structures, the $O(N^2)$ causal set generation process, and several implementations of the $O(N^3)$ algorithm to compute the Benincasa-Dowker action of compact regions of spacetime. We show that by tailoring the data structures and algorithms to take advantage of low-level CPU and GPU architecture designs, we are able to increase the efficiency and reduce the amount of required memory significantly. The presented algorithms and their implementations rely on methods that use CUDA, OpenMP, x86 Assembly, SSE/AVX, Pthreads, and MPI. We also analyze the scaling of the algorithms' running times with respect to the problem size and available resources, with suggestions on how to modify the code for future hardware architectures.
\end{abstract}


\end{frontmatter}

\noindent
{\bf PROGRAM SUMMARY} \\
\begin{small}
{\em Program Title:} Causal Set Generator and Action Computer \\
{\em URL:} \url{https://bitbucket.org/dk-lab/causalsetgenerator} \\
{\em Licensing Provisions:} MIT \\
{\em Programming Language:} C++/CUDA, x86 Assembly \\
{\em Computer:} Any with Intel CPU \\
{\em Operating System:} (RedHat) Linux \\
{\em RAM:} 512 MB \\
{\em Number of Processors Used:} 112 \\
{\em Distribution Format:} Online Repository \\
{\em Classification:} 1.5, 1.9, 6.5, 23 \\
{\em Nature of Problem:} Generate causal sets and compute the Benincasa-Dowker action. \\
{\em Solution Method:} We generate causal sets sprinkled on a Lorentzian manifold by randomly sampling element coordinates using OpenMP and linking elements using CUDA. Causal sets are stored in a minimal binary representation via the {\tt FastBitset} class. We measure the action in parallel using OpenMP, SSE/AVX and x86 Assembly. When multiple computers are available, MPI and POSIX threads are also incorporated. \\
{\em Running Time:} The runtime depends on the causal set size. A typical simulation can be performed in under a minute. Scaling with respect to Amdahl's and Gustafson's Laws is analyzed in the body of the text. \\
{\em Additional Comments:} The program runs most efficiently with an Intel processor supporting AVX2 and an NVIDIA GPU with compute capability greater than or equal to 3.0. \\
\end{small}

\section{Introduction}
There exist a multitude of viable approaches to quantum gravity, among which causal set theory is perhaps the most minimalistic in terms of baseline assumptions. It is based on the hypothesis that spacetime at the Planck scale is composed of discrete ``spacetime atoms'' related by causality~\cite{bombelli1987}. These ``atoms'', hereafter called elements, possess a partial order which encodes all information about the causal structure of spacetime, while the number of these elements is proportional to the spacetime volume---``Order + Number = Geometry''~\cite{sorkin2003}. One of the first successes of the theory was the prediction of the order of magnitude of the cosmological constant long before experimental evidence~\cite{sorkin1990}, while one of the most recent significant advances was the definition of a statistical partition function for the canonical causal set ensemble \blue{$\Omega$}~\cite{surya2012} based on the Benincasa-Dowker action~\cite{benincasa2010}. This work, \blue{which examined the space of 2D orders $\Omega_{2D}\subseteq\Omega$ defined in~\cite{brightwell2008}}, provided a framework to study phase transitions and measure observables, with paths towards developing a dynamical theory of causal sets from which Einstein's equations could possibly emerge in the continuum limit. Yet the progress along this path is partly blocked on numerical limitations. Since the theory is non-local, the combination of action computation running times, $O(N^3)$, and thermalization times, $O(N^2)$, of Monte-Carlo methods used to sample causal sets from the ensemble, result in $O(N^5)$ overall running times, limiting numerical experimentation to causal set sizes $N$ of just tens of elements. \par

Here we present new fast algorithms to generate causal sets sprinkled onto a Lorentzian manifold and to compute the Benincasa-Dowker action, with an emphasis on how these algorithms are optimized by leveraging the computer's architecture and instruction pipelines. After providing a short background on causal sets and the Benincasa-Dowker action in Sections~\ref{sec:causal_sets} and~\ref{sec:bd_action}, we describe several algorithm implementations to generate causal sets in Section~\ref{sec:construction}. Section~\ref{sec:fastbitset} presents a highly optimized data structure to represent causal sets that speeds up the computation of the action, Section~\ref{sec:action}, by orders of magnitude. Section~\ref{sec:simulations} presents an analysis of algorithms' running times as functions of the causal set size and available computational resources. We conclude with a summary in Section~\ref{sec:conclusion}.

\subsection{Causal Sets}
\label{sec:causal_sets}
\begin{figure}[!t]
\centering
\includegraphics[width=\linewidth]{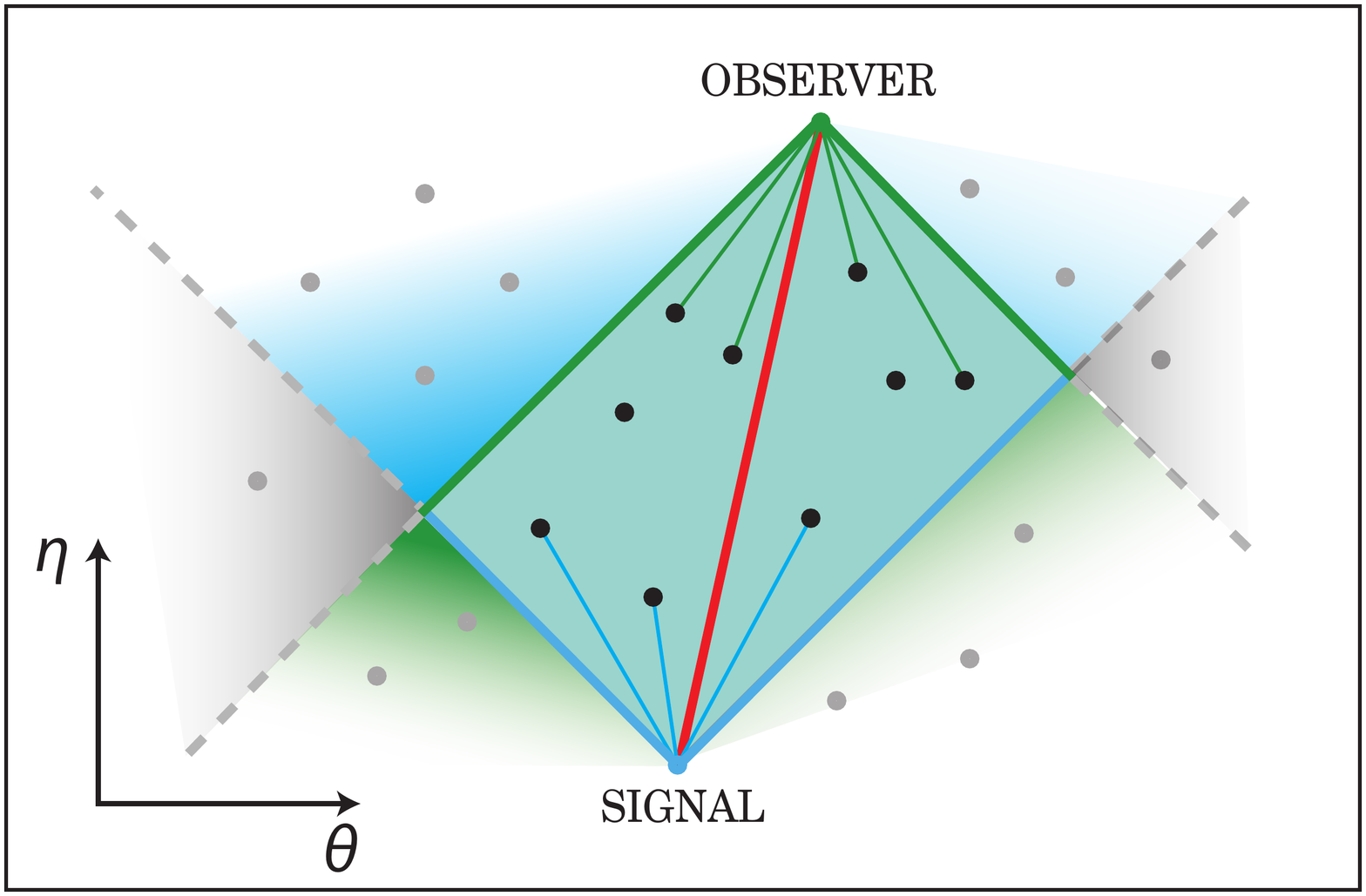}
\caption{{\bf The causal set as a random geometric graph.} \mbox{Elements} of the causal set are sprinkled uniformly at random with intensity $\xi$ into a particular region of spacetime, where $\eta$ and $\theta$ respectively refer to the temporal and spatial coordinates in $(1+1)$ dimensions. Light cones, drawn by 45-degree lines in these conformal coordinates, bound the causal future and past of each element. When light cones of a pair of elements (shown in blue and green) overlap, the elements are said to be causally related, or timelike separated, as indicated by the bold red line. The black elements both to the future of the signal and to the past of the observer form the pair's Alexandroff set shown by the teal color. Not all pairwise relations are drawn.}
\label{fig:alexandroff}
\end{figure}

Causal sets, or locally-finite partially ordered sets, are the central object in the causal set approach to quantum gravity~\cite{bombelli1987,wallden2010,surya2011}. These structures are modeled as directed acyclic graphs (DAGs) with $N$ labeled elements $(n_1,n_2,\ldots,n_N)$ and directed pairwise relations $(n_i,n_j)$. If obtained by sprinkling onto a Lorentzian manifold, they approximate the manifold in the continuum limit $N\to\infty$. Lorentzian manifolds are $(d+1)$-dimensional manifolds with $d$ spatial dimensions and one temporal dimension whose metric tensors $g_{\mu\nu}$, $\mu,\nu=0,1,\ldots,d$, have one negative eigenvalue~\cite{hawking1976,malament1977}. These DAGs are a particular type of random geometric graph~\cite{penrose2003}: elements are assigned coordinates in time and $d$-dimensional space via a Poisson point process with intensity $\xi$, and they are linked pairwise if they are causally related, i.e., timelike-separated in the spacetime with respect to the underlying metric (Figure~\ref{fig:alexandroff}). \blue{As a side note, sprinkling onto a given Lorentzian manifold is definitely not the only way to generate random causal sets. The general definition of a causal set can be found in~\cite{bombelli1987}, and random causal sets also can be obtained by sampling from the canonical ensemble $\Omega$~\cite{surya2012}, or more generally, from the ensemble of random partial orders $P_{n,p}$~\cite{winkler1985}, i.e., they can in general be treated as unlabeled partial orders.} Due to the non-locality implied by the causal structure, causal sets have an information content which scales at least as $O(N^2)$ compared to that in competing theories of discrete spacetime which scales as $O(N)$~\cite{glaser2017,surya2017,surya2017pi}. As a result, by using the causal structure information contained in these DAG ensembles, one can recover the spacetime dimension~\cite{myrheim1978,meyer1989}, continuum geodesic distance~\cite{rideout2009}, differential structure~\cite{dowker2013,glaser2014,aslanbeigi2014,belenchia2016}, Ricci curvature~\cite{benincasa2010}, and the Einstein-Hilbert action~\cite{benincasa2011,benincasa2013,buck2015,glaser2017}, among other properties.

\subsection{The Benincasa-Dowker Action}
\label{sec:bd_action}

\begin{figure*}[!t]
\centering
\includegraphics[width=0.5\textwidth]{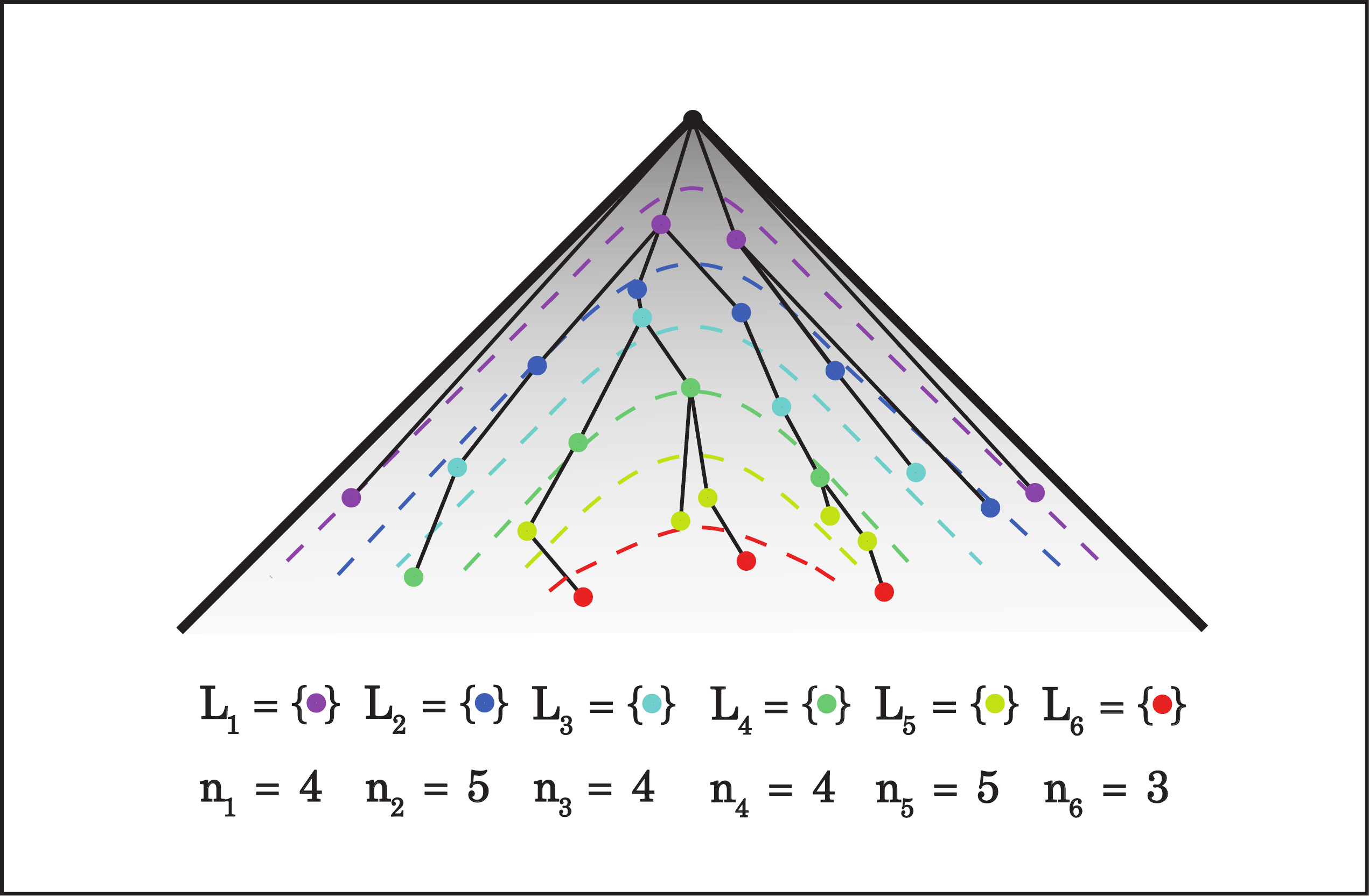}%
\includegraphics[width=0.5\textwidth]{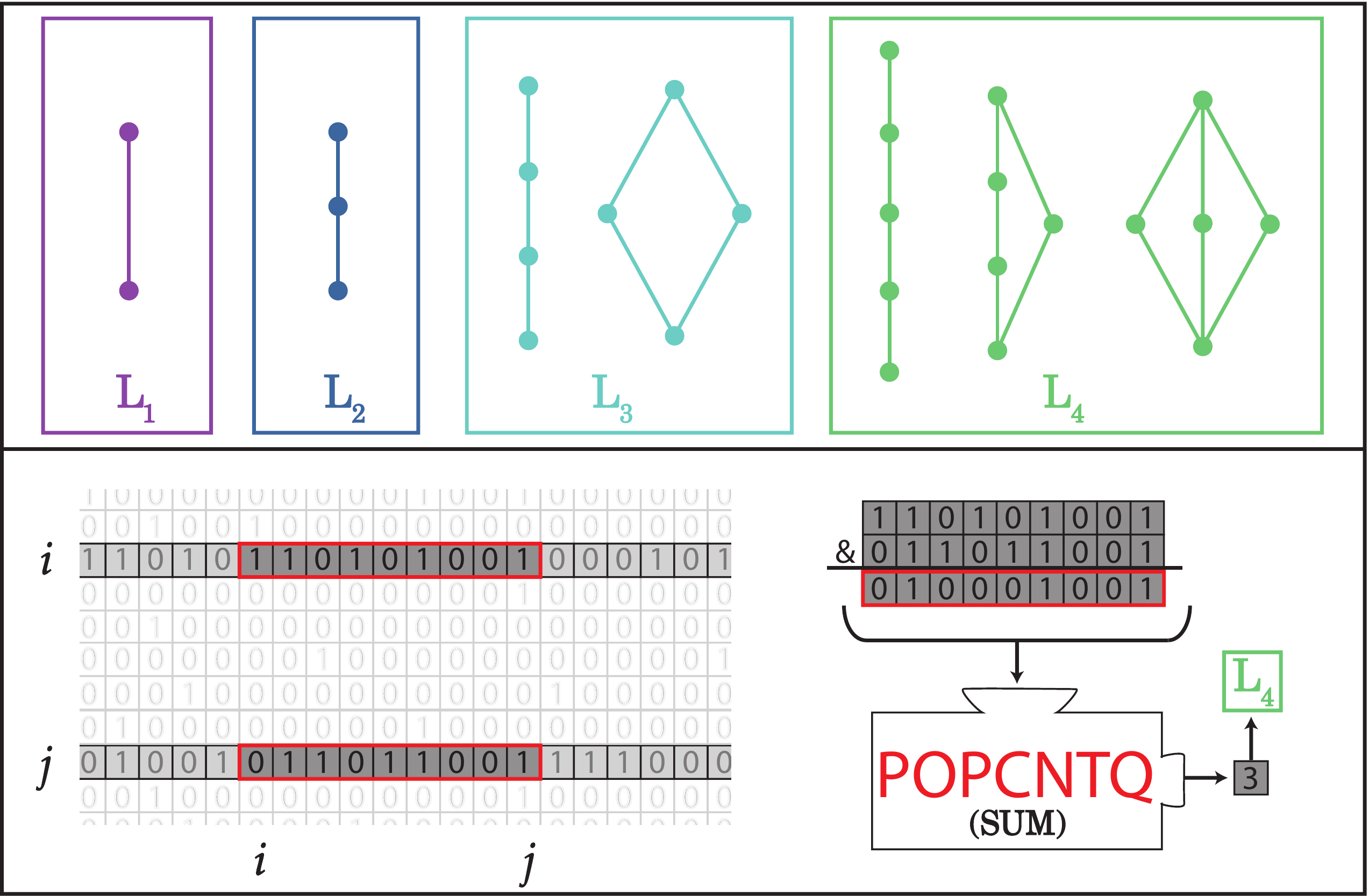}
\caption{{\bf Proper distance and the order intervals.} The left panel shows discrete hypersurfaces of constant proper time $\tau=x^2-t^2$ (dashed) are approximated using the graph distance. If the black point is some element $x$ in a larger causal set, then the order intervals would be found by counting the number of elements belonging to each hypersurface, i.e., $n_i = |L_i|$. In general the structure is not tree-like. The top of the right panel shows the subgraphs associated with each of the first four inclusive order intervals used in~\eqref{eq:s_local}, and the bottom part shows how they are detected using the causal (adjacency) matrix, assuming the graph has been topologically sorted, i.e., time-ordered. For each pair of timelike separated elements $(i,j)$, we take the inner product of rows $i$ and $j$ between columns $i$ and $j$ using the bitwise {\tt AND} in place of multiplication and the {\tt popcntq} instruction in place of a sum. The resulting value tells how many elements lie within the Alexandroff set. Details of the algorithm can be found in Section~\ref{sec:vecprod}.}
\label{fig:intervals}
\end{figure*}
In many areas of physics, the action ($S$) plays the most fundamental role: using the least action principle~\cite{maupertuis1744,gelfand1963}, one can recover the dynamic laws of the theory as the Euler-Lagrange equations that represent the necessary condition for action extremization $\delta S = 0$. In general relativity, from the Einstein-Hilbert (EH) action,
\begin{equation}
\label{eq:eh_action}
S_{EH} = \frac{1}{2}\int\!R\left(x^\mu\right)\sqrt{-g}\,dx^\mu\,,
\end{equation}
where $R$ is the Ricci scalar curvature and $g$ is the metric tensor determinant, Einstein's field equations can be explicitly derived and then solved given a particular set of constraints~\cite{wald1984}. Therefore, if one hopes to develop a dynamical theory of quantum gravity, one would hope that \blue{either} the discrete action in the quantum theory \blue{converges} to~\eqref{eq:eh_action} in the \blue{large-$N$} limit, \blue{as we find with the Regge action for gravitation~\cite{regge1961}, or an interacting theory leads to an effective action, as we see with the Wilson action in quantum chromodynamics~\cite{wilson1974}.} The numerical investigation of whether such \blue{a transition} does indeed take place can be quite difficult: the quantum gravity scale is the Planck scale, so that if the convergence is slow, it may be extremely challenging to observe it numerically. This is indeed the case for the causal set discrete action, known as the Benincasa-Dowker (BD) action~\cite{benincasa2010}, which has been shown to converge slowly to the EH action in curved higher-dimensional spacetimes such as $(3+1)$-dimensional de Sitter spacetime~\cite{benincasa2013,belenchia2016}. \par
The BD action was discovered in the study of the discrete d'Alembertian ($B$), i.e., the discrete covariant second-derivative approximating $\Box\equiv-\partial_t^2+\nabla^2$, defined in $(1+1)$ dimensions, for instance, as
\begin{equation}
\label{eq:b_local}
\begin{aligned}
B\phi\left(x^\mu\right)=\frac{2}{l^2}\Biggl(&-\phi\left(x^\mu\right)+ \\
&2\left[\sum_{y\in L_1}-2\sum_{y\in L_2}+\sum_{y\in L_3}\right]\phi\left(y^\mu\right)\Biggr)\,,
\end{aligned}
\end{equation}
where $\phi(x^\mu)$ is a scalar field on the causal set, $l\equiv\xi^{-1/(d+1)}$ is the discreteness scale, and the $i^{th}$ order inclusive order interval (IOI) $L_i$ corresponds to the set of elements $\{y\}$ which precede $x$ with exactly $(i-1)$ elements $\{z_j\}$ within each open Alexandroff set, i.e., $y\prec \{z_j\}\prec x\,\forall\,y\in L_i$ and $|\{z_j\}|=i-1$. In~\cite{benincasa2010} it was shown that in the continuum limit, \eqref{eq:b_local} converges in expectation to the continuum d'Alembertian plus another term proportional to the Ricci scalar curvature
\begin{equation}
\label{eq:dalembertian}
\lim_{N\to\infty}\mathbb{E}\left[B\phi\left(x^\mu\right)\right] = \Box\phi\left(x^\mu\right)-\frac{1}{2}R\left(x^\mu\right)\phi\left(x^\mu\right)\,.
\end{equation}
\noindent From~\eqref{eq:b_local} and~\eqref{eq:dalembertian} one can see when the field is constant everywhere, so that $\Box\phi(x^\mu)=0$, then~\eqref{eq:b_local} converges to the Ricci curvature in the continuum limit, and therefore to the EH action when summed over the entire causal set. It was also shown in~\cite{benincasa2010} that the expression for the BD action in $(1+1)$ dimensions is
\begin{equation}
\label{eq:s_local}
S_{BD} = 2(N-2n_1+4n_2-2n_3)\,,
\end{equation}
where $n_i$ is the abundance of the $i^{th}$ order IOI, i.e., the cardinality of the set $L_i$ (Figure~\ref{fig:intervals}). While~\eqref{eq:s_local} converges in expectation, any typical causal set tends to have a BD action far from the mean. This poses a serious problem for numerical experiments which already require large graphs, $N\gtrsim2^{16}$, to show convergence, and also indicates that Monte Carlo experiments must have relatively large thermalization times. To partially alleviate this problem, it is not~\eqref{eq:s_local} which one usually calculates, but rather another expression, called the ``smeared'' or ``non-local'' action ($S_\varepsilon$), which is obtained by averaging (or smearing) over subgraphs described by a mesoscale characterized by $\varepsilon\in(0,1)$. The new expression which replaces~\eqref{eq:s_local} is
\begin{equation}\label{eq:s_smeared}
\begin{split}
S_\varepsilon&= 2\varepsilon\left[N-2\varepsilon\sum_{i=1}^{N-1}n_if_2\left(i-1,\varepsilon\right)\right]\,, \\
f_2\left(i,\varepsilon\right) &= \left(1-\varepsilon\right)^i\left[1-\frac{2\varepsilon i}{1-\varepsilon}+\frac{\varepsilon^2 i\left(i-1\right)}{2\left(1-\varepsilon\right)^2}\right]\,.
\end{split}
\end{equation}
The smeared action~\eqref{eq:s_smeared} was shown to also converge to the EH action in expectation, while fluctuations are greatly suppressed so that numerical experiments with the same
degree of convergence accuracy can be performed with orders of magnitude smaller graph sizes~\cite{belenchia2016}. \par

While in some cases one might want to compare directly the expectation of the BD action to the continuum result~\eqref{eq:eh_action}, in Monte Carlo experiments with the canonical causal set ensemble one uses~\eqref{eq:s_smeared} in the quantum partition function
\begin{equation}
Z(N,d,\mathcal{T})=\sum_{C}e^{iS_\varepsilon/\hbar}\,,
\end{equation}
where the sum is over the ensemble of all causal sets $C$ with fixed size $N$, dimension $d$, and topology $\mathcal{T}$. The analytically continued partition function used in numerical experiment is
\begin{equation}
Z(N,d,\mathcal{T})=\sum_{C}e^{-\beta S_\varepsilon/\hbar}\,,
\end{equation}
where $\hbar\to1$ and $\beta\in\mathbb{R}^+$. Methods for generating causal set Markov chains using this partition function are discussed in~\blue{\cite{surya2012,henson2017}}. 

\subsection{Computational Tasks}
Generating causal sets involves an $O(N)$ coordinate generation operation followed by an $O(N^2)$ element linking operation, both of which can be parallelized (Section~\ref{sec:construction}). Yet the bottleneck is not graph generation but the $O(N^3)$ action computation. After each causal set is constructed, the primary computationally intensive task in computing~\eqref{eq:s_smeared} is counting the IOIs. For each pair of causally related elements we must count the number of elements within their Alexandroff set. As a result, the runtime depends greatly on the ordering fraction, defined as the fraction of related pairs, which in turn depends on the choice of manifold, dimension, and bounding region. \par

Previous work implemented as a part of the Cactus Framework~\cite{allen2010} has been quite successful, but because the causal set toolkit is part of a broader numerical relativity package it is challenging to modify core data structures and to take advantage of platform-specific architectures. Therefore, one of the main new features of the software suite presented here is a new efficient data structure called the {\tt FastBitset} (Section~\ref{sec:fastbitset}), which offers compressed-bit storage and several highly optimized algorithms designed specially to calculate the smeared BD action. As a result, larger causal sets may be studied in the asymptotic regime $N\gtrsim2^{16}$, possibly up to the extreme sizes $N\sim2^{24}$, and the Markov chains generated by smaller causal sets may be extended further than before to enable a closer examination of phase transitions~\blue{\cite{surya2012,glaser2017}}. \par 

\blue{We note that if other possible forms of the causal set action arise in the future, as soon as their definitions rely only on the adjacency matrix of a causal set, they can also take advantage of the presented algorithms, since these algorithms use only causal set adjacency matrices, and rely on optimized set and counting operations. For the same reasons, i.e., since these algorithms use causal set adjacency matrices only, they can be applied without modification not only to causal sets obtained by sprinkling onto a Lorentzian manifold, but also to any other causal sets, e.g., to Kleitman-Rothschild partial orders~\cite{kleitman1975}.} 

\section{Causal Set Generation}
\label{sec:construction}
\subsection{Coordinate Generation}
For a finite region of a particular Lorentzian manifold, coordinates are sampled via a Poisson point process with intensity $\xi$, using the normalized distributions given by the volume form of the metric.  For instance, for any $(d+1)$-dimensional Friedmann-Lema\^itre-Robertson-Walker (FLRW) spacetime~\cite{griffiths2009} with compact spatial hypersurfaces, the volume form may be written
\begin{equation}
dV=a(t)^ddt\,d\Omega_d\,,
\end{equation}
where $a(t)$ is the scale factor, which describes how space expands with time, and $d\Omega_d$ is the differential form for the $d$-dimensional sphere. From this expression, we find the normalized temporal distribution is $\rho(t)=a(t)^d/\int a(t^\prime)^d\,dt^\prime$, and spatial coordinates are sampled from the surface of the $d$-dimensional unit sphere. Because the $(d+1)\times N$ coordinates of the elements sprinkled within a spacetime are all independent with respect to each other, these may easily be generated in parallel using OpenMP, \blue{which is a C/C++ and Fortran library used to distribute parallel tasks over multiple CPU cores~\cite{openmp}}.

\subsection{Pairwise Relations}
Once coordinates are assigned to the elements, the pairwise relations are found by identifying pairs of elements which are timelike separated, and efficient storage requires the proper choice of the representative data structure. A causal set is a graph, i.e., a set of $N$ labeled elements along with a set of pairs $(i,j)$ which describe pairwise relations between elements, so the most straightforward representation uses an adjacency matrix of size $N\times N$. If the graph is simply-connected, i.e., there exist no self-loops or multiply-connected pairs, then this matrix contains only 1's and 0's, with each entry indicating the existence or non-existence of a relation between the pair of elements specified by a particular pair of row and column indices. Moreover, if this graph is undirected, the matrix will be symmetric. We represent \blue{naturally ordered} causal sets as undirected graphs with topologically sorted elements, meaning that elements are labeled such that an element with a larger index will never precede an element with a smaller index. In the context of a conformally flat embedding space, \blue{which is the only type we consider in this work,} this simply means elements are sorted by their time coordinate before relations are identified. \blue{Yet this does not mean that the presented causal set generation algorithms are impossible to adjust to generate causal set sprinkled onto spacetimes that are not conformally flat. Indeed, in such spacetimes topological sorting can be used, as any partial order can be topologically sorted by the order-extension principle~\cite{brightwell1991}.}

\subsubsection{Naive CPU Linking Algorithm}
The naive implementation of the linking algorithm using the CPU uses a sparse representation in the compressed sparse row format~\blue{\cite{sato1963,tinney1967}}. Because the elements have been sorted, we require twice the memory to store sorted lists of both future-directed and past-directed relations, i.e., one list identifies relations to the future and the other those to the past. While identification of the relations is in fact only $O(N^2)$ in time, the data reformatting (list sorting) pushes it roughly to $O(N^{2.6})$, as we will see in Section~\ref{sec:simulations}. 

\subsubsection{OpenMP Linking Algorithm}
The second implementation uses the dense graph representation and is parallelized using OpenMP. Using this dense representation for a sparse graph can waste a relatively large amount of memory compared to the information content; however, the nature of the problem described in the previous section dictates a dense representation will permit a much faster algorithm, as we will discuss later in Sections~\ref{sec:action} and~\ref{sec:simulations}. Moreover, the sparsity will depend greatly on the input parameters, so in many cases the binary adjacency matrix is the ideal representation.

\subsubsection{Naive GPU Linking Algorithm}
While OpenMP offers a great speedup over the naive implementation, the linking algorithm is several orders of magnitude faster when instead we use one or more Graphics Processing Units (GPUs) with the CUDA library~\cite{cuda}. \blue{Since they have many more cores than CPUs, GPUs are typically best at solving problems which require many thousands of independent low-memory tasks to be performed.} There are many difficulties in designing appropriate algorithms to run on a GPU: one must consider size limitations of the global memory, which is the GPU equivalent of the RAM, and the GPU's L1 and L2 \blue{memory} caches, as well as the most efficient memory access patterns. One particularly common optimization uses the shared memory, which is a reserved portion of up to 48 KB of the GPU's 64 KB L1 cache. This allows a single memory transfer from global memory to the L1 cache so that spatially local memory reads and writes by individual threads afterwards are at least 10x faster. At the same time, an additional layer of synchronizations among threads in the same thread block \blue{(i.e., threads which execute concurrently)} must be considered to avoid thread divergence~\blue{\cite{wong2010}} and unnecessary \blue{\texttt{if/else}} branching. It also puts constraints on data structures since it requires spatially local data or else the cache miss rate, \blue{i.e., the percent of time data is pulled from the RAM instead of the cache}, will drastically increase. \par

The first GPU implementation offers a significant speedup by allowing each of the 2496 cores in the NVIDIA K80m (using a single GK210 processor) to perform a single comparison of two elements. The output is a sparse edge list of 64-bit unsigned integers, so that the lower and upper 32 bits each contain a 32-bit unsigned integer corresponding to a pair of indices of related elements. After the list is fully generated, it is decoded on the GPU using a parallel bitonic sort to construct the past and future sparse edge lists. During this procedure, vectors containing degree data are also constructed by counting the number of writes to the edge list.

\subsubsection{Optimized GPU Linking Algorithm}
Despite the great increase in efficiency, this method fails if $N$ is too large for the edge list to fit in global GPU memory or if $N$ is not a multiple of 256. The latter failure occurs because the thread block size is set to 128 for architectural reasons~\footnote{On the NVIDIA K80m, which has a Compute Capability of 3.7, each thread block cannot have greater than 1024 threads, there can be at most 16 thread blocks per multiprocessor, and at the same time no greater than 2048 threads per multiprocessor.}, and the factor of two comes from the index mapping used internally which treats the adjacency matrix as four square submatrices of equal size. The second GPU implementation addresses these limitations by tiling the adjacency matrix, i.e., sending smaller submatrices to the GPU serially.  Further, when $N$ is not a round number these edge cases are handled by exiting threads with indices outside the proper bounds so that no improper memory accesses are performed. \par

This second implementation also greatly improves the speed by having each thread work on four pairs of elements instead of just one. Since each of the four pairs has the same first element by construction, the corresponding data for that element may be read into the shared memory, thereby reducing the number of accesses to global memory. Moreover, threads in the same thread block also use shared memory for the second element in each pair. Hence, since each thread block has 128 threads and each thread works on four pairs, there are only 132 reads (128+4) to global memory rather than 512 (128$\times$4), where each read consists of reading $(d+1)$ floats for a $(d+1)$-dimensional causal set. Finally, when the dense graph representation is used, the decoding step may be skipped, which offers a rather substantial speedup when the graph is dense. There are other optimizations to reduce the number of writes to global memory using similar techniques via the shared memory cache.

\subsubsection{Asynchronous GPU Linking Algorithm}
A third version of the GPU linking algorithm also exists which uses asynchronous CUDA calls to multiple concurrent streams~\blue{\cite{cuda}}. By further tiling the problem, simultaneously data can be passed to and from the GPU while another stream executes the kernel, i.e., the linking operations. This helps reduce the required bandwidth over the PCIe bus, \blue{which connects the GPU to the CPU and other devices,} and can sometimes improve performance when the data transfer time is on par with the kernel execution time. We find in Section~\ref{sec:simulations} this does not provide as great a speedup as we expected, so this is one area for future improvement should this end up being a bottleneck in other applications.

\section{The {\tt FastBitset} Class}
\label{sec:fastbitset}
\subsection{Problems with Existing Data Structures}
The relations found by the linking algorithm are best stored in dense matrix format for the action algorithm, as we will see in Section~\ref{sec:action}. A binary adjacency matrix can be implemented in several ways in C++. The naive approach is to use a {\tt std::vector<bool>} object. While this is a compact data structure, there is no guarantee memory is contiguously stored internally and, moreover, reading from and writing to individual locations is computationally expensive. Because the data is stored in binary, there is necessarily an internal conversion involving several bitwise and type-casting operations which make these simple operations take longer than they would for other data structures. \par

The next best option is the {\tt std::bitset<>} object. This is a better option than the {\tt std::vector<bool>} because it has bitwise operators pre-defined for the object as a whole, i.e., to multiply two objects one need not use a {\tt for} loop; rather, operations like {\tt c = a \& b} are already implemented. Further, it has a bit-counting operation defined, making it easy to immediately count the number of bits set to `1' in the object. Still, there is no guarantee of contiguous memory storage and, worst of all, the size must be known at compile-time. These two limitations make this data structure impossible to use if we want to specify the size of the causal set at runtime. \par

Finally, the last option we'll examine is the {\tt boost::dynamic\_bitset<>} provided in the Boost C++ Libraries~\cite{boost}. While this is not a part of the ISO C++ Standard, it is a well-maintained and trusted library. Boost is known for offering more efficient implementations of many common data structures and algorithms. The {\tt boost::dynamic\_bitset<>} can be dynamically sized, unlike the {\tt std::bitset<>}, the memory is stored contiguously, and it even has pre-defined bitwise and bit-counting operations. Still, it does not suit the needs of the abovementioned problem because it is not possible to access individual portions of the bitset: we are limited to work only with individual bits or the entire bitset. \par

Given these limitations, we have developed the {\tt FastBitset} class to represent causal sets in a way which is most efficient for non-local algorithms such as the one used to find the BD action. The adjacency matrix is comprised of a {\tt std::vector} of these {\tt FastBitset} objects, with each object corresponding to a row of the matrix. Internally, this data structure holds an array of 64-bit unsigned integers, referred to as blocks, which contain the matrix elements in their raw bits. We have provided all four set operations (intersection, union, disjoint union, and difference) and several bit-counting operations, including variations which maybe used on a proper subset of the entire object. The performance-critical algorithms used to calculate the BD action have been optimized using inline assembly and \blue{Intel's Streaming SIMD Extensions (SSE) and Advanced Vector Extensions (AVX)} instructions~\cite{intel}. \par

\subsection{Optimized Algorithms in the {\tt FastBitset}}
One of the most frequently used operations \blue{in the action calculation} is the set intersection, i.e., row multiplication using the bitwise {\tt AND} operator \blue{(Figure~\ref{fig:intervals}(right))}. The naive implementation uses a {\tt for} loop, but the optimized algorithm takes advantage of the 256-bit YMM registers located within each physical CPU core~\blue{\cite{intel}. For a review of x86 microarchitectures, see~\cite{weidendorfer2011,fog2017microarchitecture}.} The larger width of these registers means that in a single CPU cycle we may perform a bitwise {\tt AND} on four times the number of bits as in the naive implementation at the expense of moving data to and from these registers. The outline is described in Algorithm~\ref{alg:intersection}. It is important to note that for such an operation to be possible, the array of blocks must be 256-bit aligned. Any bits used as padding are always set to zero so they do not affect any results.

\begin{algorithm}
\caption{Set Intersection with AVX}
\label{alg:intersection}
\begin{algorithmic}[1]
\Input
\Statex $A$ \Comment The bit array of the first {\tt FastBitset}
\Statex $B$ \Comment The bit array of the second {\tt FastBitset}
\Statex $n$ \Comment The number of blocks

\Procedure{intersection}{A,B,n}
\For { $i=0;~i < n;~i\pluseq4$ }
\State {\tt ymm0} $\gets A[i]$
\State {\tt ymm1} $\gets B[i]$
\State {\tt ymm0} $\gets$ ({\tt ymm0}) \& ({\tt ymm1}) \label{op:vpand}
\State $A[i] \gets$ {\tt ymm0}
\EndFor
\EndProcedure

\Output
\Statex $A$ \Comment The first bit array now holds the result
\end{algorithmic}
\end{algorithm}

The code shown inside the {\tt for} loop is written entirely in inline assembly, with Operation~\ref{op:vpand} using the SIMD instruction {\tt vpand} provided by AVX. Therefore, for each set of 256 bits, we use two move operations from the L1 or L2 cache to the YMM registers, one bitwise {\tt AND} operation, and one final move operation of the result back to the general purpose registers. The bottleneck in this operation is not the bitwise operation, but rather the move instructions {\tt vmovdqu}, which limits throughput due to the bus bandwidth to these registers. As a result, it is not faster to use all 16 of the YMM registers, but rather only two. While certain prefetch instructions were tested we found no further speedup. \par

One of the reasons this data structure was developed was so we could perform such an operation on a subset of two sets of bits. We apply the same principle as in Algorithm~\ref{alg:intersection}, but with unwanted bits masked out, i.e., set to zero after the operation. For blocks which lie outside the range we want to study, they are not even included in the {\tt for} loop. The new operation, denoted the {\it partial intersection}, is outlined in Algorithm~\ref{alg:partial_intersection}.

\begin{algorithm}
\caption{Partial Intersection with AVX}
\label{alg:partial_intersection}
\begin{algorithmic}[1]
\Input
\Statex $A$ \Comment The first bit array
\Statex $B$ \Comment The second bit array
\Statex $o$ \Comment Starting bit index
\Statex $n$ \Comment Length of subset
\Function {get\_bitmask}{offset}
\State \Return $(1\ll\mathrm{offset})-1$
\EndFunction

\Procedure{partial\_intersection}{A,B,o,n}
\LineComment Divide $o$ by 64 to get the block index
\State $x\gets o/64$
\LineComment Indices within the blocks
\State $a\gets o \mathmod 64$
\State $b\gets (o+n) \mathmod 64$
\If {range inside single block}
\parState {$A[x]\gets A[x] \mathand B[x] \mathand$ {\tt get\_bitmask($a$)} $\mathand$ {\tt get\_bitmask($b$)}}
\State $u\gets 1$ \Comment Used one block
\Else
\LineComment{Intersection on full blocks}
\State $m\gets(n-1)/64$ \Comment Number of full blocks
\State {\tt intersection($A[x+1], B[x+1], m$)}
\LineComment {Intersection on end blocks}
\State $A[x] \andeq B[x] \mathand$ {\tt get\_bitmask($a$)}
\State $A[x+m] \andeq B[x+m] \mathand$ {\tt get\_bitmask($b$)}
\State $u\gets m+2$ \Comment Used $m+2$ blocks
\EndIf

\LineComment {Set other blocks to zero}
\State $l\gets a$
\State $h\gets$ {\tt $A$.getNumBlocks()}$-l-u$
\If {$l > 0$}
\State {\tt memset($A,0,8*l$)}
\EndIf
\If {$h > 0$}
\State {\tt memset($A[l+u],0,8*h$)}
\EndIf
\EndProcedure

\Output
\Statex $A$ \Comment The first bit array now holds the result
\end{algorithmic}
\end{algorithm}

\begin{algorithm}
\caption{Optimized Bit Counting}
\label{alg:popcnt}
\begin{algorithmic}[1]
\Input
\Statex $A$ \Comment The bit array
\Statex $N$ \Comment The number of blocks

\Procedure{count\_bits}{A,n}
\LineComment {The counter variables}
\State $c[4]\gets \{0,0,0,0\}$
\For {{$i = 0; i < N; i \pluseq 4$}}
\State $A[i]\gets ${\tt popcntq($A[i]$)}
\State $c[0]\pluseq A[i]$
\State $A[i+1]\gets ${\tt popcntq($A[i+1]$)}
\State $c[1]\pluseq A[i+1]$
\State $A[i+2]\gets ${\tt popcntq($A[i+2]$)}
\State $c[2]\pluseq A[i+2]$
\State $A[i+3]\gets ${\tt popcntq($A[i+3]$)}
\State $c[3]\pluseq A[i+3]$
\EndFor
\EndProcedure

\Output
\Statex $c[0]+c[1]+c[2]+c[3]$ \Comment Number of set bits
\end{algorithmic}
\end{algorithm}

In the partial intersection algorithm, we consider two scenarios: in one the entire range of bits lies within a single block, and in the second it lies over some range of blocks, in which case the original intersection algorithm may be used on those full blocks. In either case, it is essential all bits outside the range of interest are set to zero, as shown by the {\tt memset} and {\tt get\_bitmask} operations. \par

The final operation which we must optimize \blue{to efficiently calculate the action} is the bit count and, therefore, the partial bit count as well. This is a well-studied operation which has many implementations and is strongly dependent on the hardware and compiler being used. The bit count operation takes some binary string, usually in the form of an unsigned integer, and returns the number of bits set to one. Because it is such a fundamental operation, some processors support a native assembly instruction called {\tt popcnt} which acts on a 32- or 64-bit  unsigned integer. Even on systems which support these instructions, the compiler is not always guaranteed to choose these instructions. For instance, the GNU function {\tt \_\_builtin\_popcount} actually uses a lookup table, as does Boost's {\tt do\_count} method used in its {\tt dynamic\_bitset}. Both are rather fast, but they are not fully optimized, and for this reason we will attempt to package the fastest known implementation with the {\tt FastBitset}. When such an instruction is not supported the code will default to Boost's implementation. \par

The fastest known implementation of the bit count algorithm uses the native 64-bit CPU instruction {\tt popcntq}, where the trailing `q' indicates the instruction operates on a (64-bit) quadword operand. While we could use a {\tt for} loop with a simple assembly call, we would not be taking advantage of the modern pipeline architecture~\blue{\cite{fog2017microarchitecture}} with just one call to one register. For this reason, we unroll the loop and perform the operation in pseudo-parallel fashion, i.e., in a way in which prefetching and prediction mechanisms will improve the instruction throughput by our explicit suggestions to the out-of-order execution (OoOE) units in the CPU. We demonstrate how this works in Algorithm~\ref{alg:popcnt}. \par

This algorithm is so successful because the instructions are not blocked nearly as much here as if they were performed using a single register. This is because the {\tt popcnt} instruction has a latency of three cycles, but a throughput of just one cycle, meaning $x$ {\tt popcnt} instructions can be executed in $x+2$ cycles instead of $3x$ cycles when they are all independent operations~\cite{fog2017instruction}. As a result, the Intel instruction pipeline allows the four sets of operations to be performed nearly simultaneously (i.e., instruction-level parallelism) via the OoOE units. While it would be possible to extend this performance to use another four registers, this would then mean the bitset would need to be 512-bit aligned. \par

\subsection{The Vector Product}
\label{sec:vecprod}
To execute the vector product operation, we want to utilize the features described above. If the {\tt popcnt} is performed directly after the intersection, a lot of time is wasted copying data to and from YMM registers when the sum variable could be stored directly in the YMM registers, for instance. Since the {\tt vmovdqu} operations are comparatively expensive, removing one out of three offers a great speedup. Furthermore, for large bitsets it is in fact faster to use an AVX implementation of the bit count~\cite{mula2017}. We show such an implementation below in Algorithm~\ref{alg:vecprod}.

\begin{algorithm}
\caption{Optimized Vector Product}
\label{alg:vecprod}
\begin{algorithmic}[1]
\Input
\Statex $A$ \Comment The first bit array
\Statex $B$ \Comment The second bit array
\Statex $N$ \Comment The number of blocks

\Procedure{vecprod}{A,B,N}
\State {\tt ymm2}$\gets${\tt table} \Comment Lookup table
\State {\tt ymm3}$\gets${\tt 0xf} \Comment Mask variable
\For {{$i=0; i<N; i\plusplus$}}
\State {\tt ymm0}$\gets A[i]$
\State {\tt ymm1}$\gets B[i]$
\State {\tt ymm0}$\gets${\tt (ymm0) \& (ymm1)} \Comment Intersection
\State {\tt ymm4}$\gets${\tt (ymm0) \& (ymm3)} \Comment Lower Mask
\State {\tt ymm5}$\gets${\tt ((ymm0) $\gg 4$) \& (ymm3)} \Comment High Mask
\State {\tt ymm4}$\gets${\tt vpshufb(ymm2, ymm4)} \Comment Shuffle
\State {\tt ymm5}$\gets${\tt vpshufb(ymm3, ymm5)} \Comment Shuffle
\State {\tt ymm5}$\gets${\tt vpaddb(ymm4, ymm5)} \Comment Horiz. Add
\State {\tt ymm5}$\gets${\tt vpsadbw(ymm5, ymm7)} \Comment Horiz. Add
\State {\tt ymm6}$\gets${\tt ymm5}$+${\tt ymm6} \Comment Accumulator
\EndFor
\State $c\gets${\tt ymm6}
\EndProcedure

\Output
\Statex $c[0]+c[1]+c[2]+c[3]$ \Comment Vector product sum
\end{algorithmic}
\end{algorithm}

This algorithm is among the best known SIMD algorithms for bit accumulation~\cite{mula2017}. At the very start, a lookup table and mask variable are each loaded into a YMM register. The table is actually the first half of the Boost lookup table, stored as an {\tt unsigned char} array. These variables are essential for the instructions later to work properly, but their contents are not particularly interesting. Once the intersection is performed, two mask variables are created using the preset mask. The bits in these masks are then shuffled ({\tt vpshufb}) according to the contents of the lookup table in a way which allows the horizontal additions ({\tt vpaddb}, {\tt vpsadbw}) to store the sum of bits in each 64-bit range in the respective range. Finally, the accumulator saves these values in {\tt ymm6}. The instructions are once again paired in a way which allows the instruction throughput to be maximized via instruction-level parallelism, and the partial vector product uses a very similar setup to the partial intersection with respect to masking and {\tt memset} operations. If the bitset is too short, i.e., if the causal set is too small, this algorithm will perform poorly due to the larger number of instructions, though it is easy to experimentally determine which to use on a particular system and then hard-code a threshold. \par

All of the algorithms mentioned so far may be easily optimized for a system with (512-bit) ZMM registers, and we should expect the greatest speedup for the set operations. Using Intel Skylake X-series and newer processors, which support 512-bit SIMD instructions, we may replace something like {\tt vpand} with the 512-bit equivalent {\tt vpandd}. An optimal configuration today would use a Xeon E3 processor with a Kaby Lake microarchitecture, which can have up to a 3.9 GHz base clock speed, together with a Xeon Phi Knights Landing co-processor, where AVX-512 instructions may be used together with OpenMP to broadcast data over 72 physical (288 logical) cores.

\section{Action Computation}
\label{sec:action}
\subsection{Naive Action Algorithm}
The optimizations described above which use AVX and OpenMP are orders of magnitude faster than the naive action algorithm, which we review here. The primary goal in the action algorithm is to identify the abundance $n_i$ of the subgraphs $L_i$ identified in Figure~\ref{fig:intervals}. When we use the smeared action rather than the local action, this series of subgraphs continues all the way up to those defined by the set of elements $L_{N-2}$, i.e., the largest possible subgraph is an open Alexandroff set containing $N-2$ elements. Therefore, the naive implementation of this algorithm is an $O(N^3)$ procedure which uses three nested {\tt for} loops to count the number of elements in the Alexandroff set of every pair of related elements. For each non-zero entry $(i,j)$ of the causal matrix, with $i<j$ due to time-ordering, we calculate the number of elements $k$ both the future of element $i$ and to the past of element $j$ and then add one to the array of interval abundances at index $k$.

\subsection{OpenMP Action Algorithm}
The most obvious optimization uses OpenMP to parallelize the two outer loops of the naive action algorithm, since the properties of each Alexandroff set in the causal set are mutually independent. Therefore, we combine the two outer loops into a single loop of size $N(N-1)/2$ which is parallelized with OpenMP, and then keep the final inner loop serialized. When we do this, we must make sure we avoid write conflicts to the interval abundance array: if two or more threads try to modify the same spot in the array, some attempts may fail. To avoid this, we generate $T$ copies of this array so that each of the $T$ threads can write to its own array. After the action algorithm has finished, we perform a reduction on the $T$ arrays to add all results to the first array in the master thread. This algorithm still scales like $O(N^3)$ since the outer loop is still $O(N^2)$ in size.

\subsection{AVX Action Algorithm}
The partial vector product algorithms described in Section~\ref{sec:vecprod} naturally provide a highly efficient modification to the naive action algorithm. The partial intersection returns a binary string where indices with 1's indicate elements both to the future of element $i$ and to the past of element $j$, and then a bit count will return the total number of elements within this interval. A summary of this procedure is given in Algorithm~\ref{alg:cardinalities}.

\begin{algorithm}
\caption{Optimized Cardinality Measurement}
\label{alg:cardinalities}
\begin{algorithmic}[1]
\Input
\Statex $A$ \Comment The adjacency matrix
\Statex $c$ \Comment The array of cardinalities
\Statex $p$ \Comment The number of element pairs

\Procedure{cardinality}{A,c,p}
\For {{$k=0; k < p; k\plusplus$}}
\LineComment {Convert the pair index to two element indices}
\State $\{i,j\}\gets${\tt convert\_index($k$)}
\If {elements are not related}
\State continue
\EndIf
\LineComment {Cardinality for pair $(i,j)$}
\State $m\gets A[i].${\tt partial\_vecprod(}$A[j],i,j-i+1${\tt )}
\State $c[m+1]\plusplus$
\EndFor
\EndProcedure

\Output
\Statex $c$ \Comment The populated array
\end{algorithmic}
\end{algorithm}

This algorithm is able to be further optimized by using OpenMP with a {\tt reduction} clause \blue{(which prevents write conflicts)} to accumulate the cardinalities. In turn, each physical core is parallelizing instructions via AVX, and then each CPU is parallelizing instructions by distributing tasks in this outer loop to each core. While it is typical to use the number of logical cores during OpenMP parallelization, we instead use the number of physical cores (typically half the logical cores, or a quarter in a Xeon Phi co-processor) because it is not always efficient to use hyperthreading alongside AVX. \par

\subsection{MPI Optimization: Static Design}
\label{sec:mpi_static}
When the graph is small, so that the entire adjacency matrix fits in memory on each computer, we can simply split the {\tt for} loop in Algorithm~\ref{alg:cardinalities} evenly among all the cores on all computers using a hybrid OpenMP and Platform MPI approach. But when the graph is extremely large, e.g., $N\gtrsim 2^{21}$, we cannot necessarily fit the entire adjacency matrix in memory. To address this limitation, we use MPI to split the entire problem among $2^x$ computers, where $x\in\mathbb{Z}^+$. Each computer will generate some fraction of the element coordinates, and after sharing them among all other computers, will generate its portion of the adjacency matrix, hereafter referred to as the adjacency submatrix. In general, these steps are fast compared to the action algorithm. \par

The MPI version of the action algorithm is performed in several steps. It begins by performing every pairwise operation possible on each adjacency submatrix, without any memory swaps among computers. Afterward, each adjacency submatrix is labeled by two numbers: the first refers to the first half of rows of the adjacency submatrix on that computer while the second corresponds to the second half, so that there are $2^{x+1}$ groups of rows labeled $\{0,\ldots,2^{x+1}-1\}$. There will never be an odd number since the matrix is 256-bit aligned. We then wish to perform the minimal number of swaps of these row groups necessary to operate on every pair of rows of the original matrix. Within each row group all pairwise operations have already been performed, so moving forward only operations among rows of different groups are performed. \par

\begin{table}[t]
\centering
\begin{tabularx}{\linewidth}{*{8}{|>{\centering\arraybackslash}X}|}
\hline
\multicolumn{2}{|c|}{Rank 0} &%
\multicolumn{2}{c|}{Rank 1} &%
\multicolumn{2}{c|}{Rank 2} &%
\multicolumn{2}{c|}{Rank 3} \\
\hline
0 & 1 & 2 & 3 & 4 & 5 & 6 & 7 \\ \hline
0 & 3 & 2 & 5 & 4 & 7 & 6 & 1 \\ \hline
0 & 5 & 2 & 7 & 4 & 1 & 6 & 3 \\ \hline
0 & 7 & 2 & 1 & 4 & 3 & 6 & 5 \\ \hline
0 & 2 & 1 & 3 & 4 & 6 & 5 & 7 \\ \hline
0 & 4 & 1 & 5 & 2 & 6 & 3 & 7 \\ \hline
0 & 6 & 1 & 7 & 4 & 2 & 5 & 3 \\
\hline
\end{tabularx}
\caption{{\bf Permutations of MPI buffers using four computers.} Each of four computers, identified by its rank, holds a quarter of the adjacency matrix. Two buffers on each computer each hold an eighth of the entire matrix, labeled $\{0,\ldots,7\}$, so that all pairwise row operations may be performed using the minimal number of inter-rank transfers. Each of the seven rows is a non-trivial permutation of the eight buffers, indicating only six rounds of MPI data transfers are necessary to calculate the action when the algorithm is split over four computers.}
\label{tbl:perms}
\end{table}

We label all possible permutations except those which provide trivial swaps, i.e., moves which would swap \blue{the submatrix rows in memory} buffers within a single computer, or moves which swap buffers in only some computers. The non-trivial configurations are shown for four computers in Table~\ref{tbl:perms}. By organizing the data in this way, we can ensure no computer will be idle after each data transfer. We use a cycle sort to determine the order of permutations so that we can use the minimal number of total buffer swaps. We are able to simulate this using a simple array of integers populated by a given permutation, after which the actual operation takes place. By starting at the current permutation and sorting to each unvisited permutation, we can record how many steps each would take. Often it is the case that several will use the same number of steps, in which case we may move from the current permutation to any of the others which use the fewest number of swaps. Once all pairwise partial vector products have completed on all computers for a particular permutation, that permutation is removed from the global list of unused permutations which is shared across all computers.
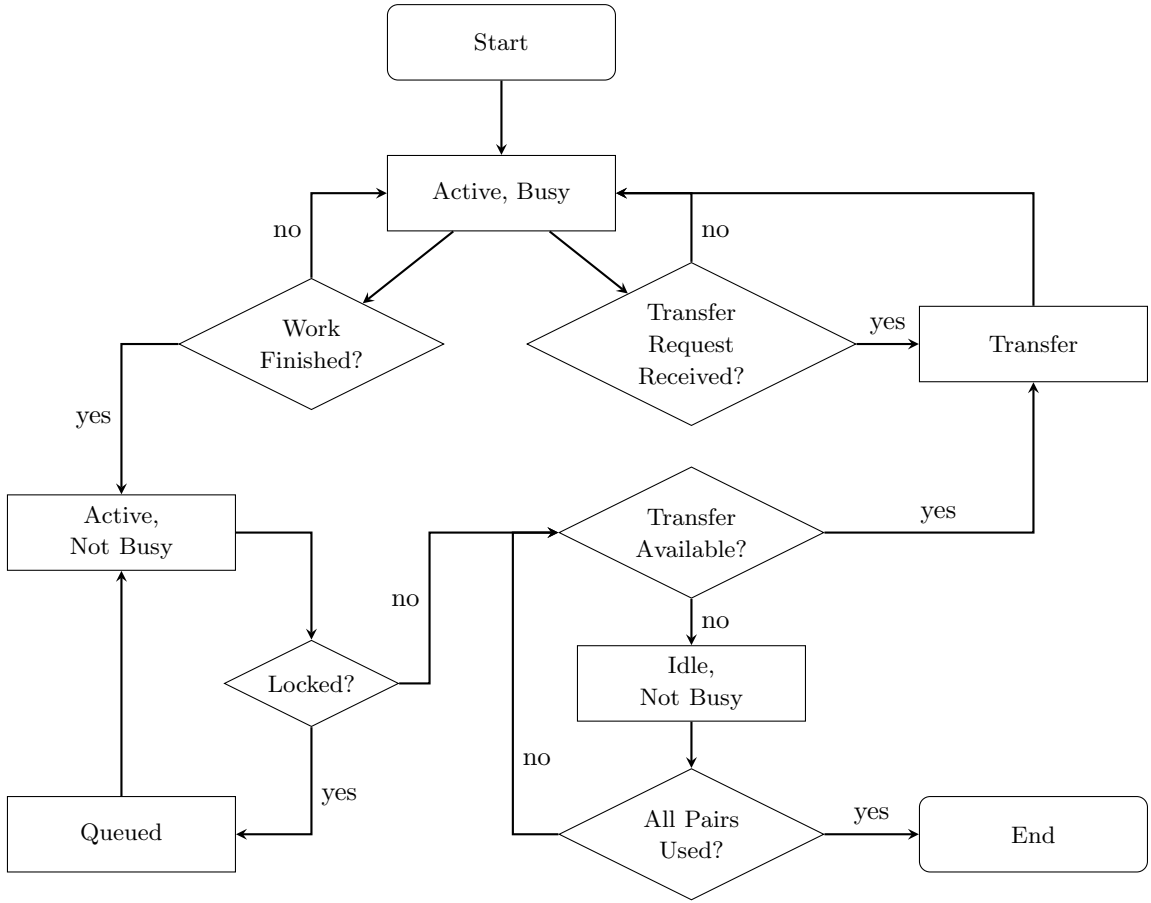
\begin{figure*}[t]
\centering
\hspace*{1.5cm}
\begin{tikzpicture}[node distance=2cm]
\node (start) [startstop] {\small Start};
\node (ab) [process, below of = start] {\small Active, Busy};
\node (wfin) [decision, below of = ab, xshift=-2.5cm, text width=1.5cm, aspect=2] {\small Work Finished?};
\node (treq) [decision, below of = ab, xshift=2.5cm, text width=1.5cm, aspect=2] {\small Transfer Request Received?};

\node (trans) [process, right of = treq, xshift=2.5cm] {\small Transfer};

\node (anb) [process, below of = wfin, text width=1.5cm, yshift=-0.5cm, xshift=-2.5cm] {\small Active, Not Busy};
\node (lok) [decision, below of = anb, aspect=2, xshift=2.5cm] {\small Locked?};
\node (que) [process, below of = lok, xshift=-2.5cm] {\small Queued};

\node (tra) [decision, right of = anb, text width=1.5cm, aspect=2, xshift=5.5cm] {\small Transfer Available?};
\node (inb) [process, below of = tra, text width=1.5cm] {\small Idle,\\ Not Busy};
\node (apu) [decision, below of = inb, text width=1.5cm, aspect=2] {\small All Pairs Used?};

\node (end) [startstop, right of = apu, xshift=2.5cm] {\small End};

\draw [arrow] (start) -- (ab);
\draw [arrow] (ab) -- (wfin);
\draw [arrow] (wfin) |- node[anchor=east, yshift=-0.5cm] {no} (ab);
\draw [arrow] (ab) -- (treq);
\draw [arrow] (treq) |- node[anchor=west, yshift=-0.5cm] {no} (ab);

\draw [arrow] (treq) -- node[anchor=south] {yes} (trans);
\draw [arrow] (trans) |- (ab);

\draw [arrow] (wfin) -| node[anchor=east, yshift=-1cm] {yes} (anb);
\draw [arrow] (anb) -| (lok);
\draw [arrow] (lok) |- node[anchor=west, yshift=0.5cm] {yes} (que);
\draw [arrow] (que) -- (anb);

\draw [arrow] (lok.east) -- +(0.4,0) |- node[anchor=east, yshift=-0.9cm] {no} (tra);
\draw [arrow] (tra) -| node[anchor=south, xshift=-1.25cm] {yes} (trans);
\draw [arrow] (tra) -- node[anchor=west] {no} (inb);
\draw [arrow] (inb) -- (apu);
\draw [arrow] (apu.west) -- +(-0.6,0) |- node[anchor=west, yshift=-3cm] {no} (tra);

\draw [arrow] (apu) -- node[anchor=south] {yes} (end);
\end{tikzpicture}
\caption{{\bf Load-balanced action algorithm using MPI.} When the adjacency matrix is split among multiple computers, we want to make sure no computers end up idle for long periods of time, yet to move from an Idle to Busy state at least one other computer must have finished its work. Initially, all computers are Active and Busy, indicating they are not waiting for another task to finish and are currently working on the action algorithm. If two other computers have requested an exchange, an Active, Busy computer will allow them to use part of its memory for temporary storage (Transfer). Once a computer finishes its portion of work on the action algorithm, it will enter the Active, Not Busy state, at which point it will add its pair of buffer indices to the global list of available buffers. An MPI spinlock, developed specifically for this algorithm, is implemented to ensure only one computer can manage a transfer. If another pair of computers is exchanging data, the Active, Not Busy computer will enter a Queued state, where it will remain until other transfers have completed. Otherwise, it will attempt a memory transfer if possible by checking the list of available buffers. If no other buffers are available, or if any available transfers would lead to redundant calculations, the computer enters the Idle, Not Busy state, where it waits for another computer to initiate a transfer. Once all buffer pairs have been used, the algorithm ends.}
\label{fig:mpi_flowchart}
\end{figure*}

\begin{figure*}[!t]
\centering
\includegraphics[width=\textwidth]{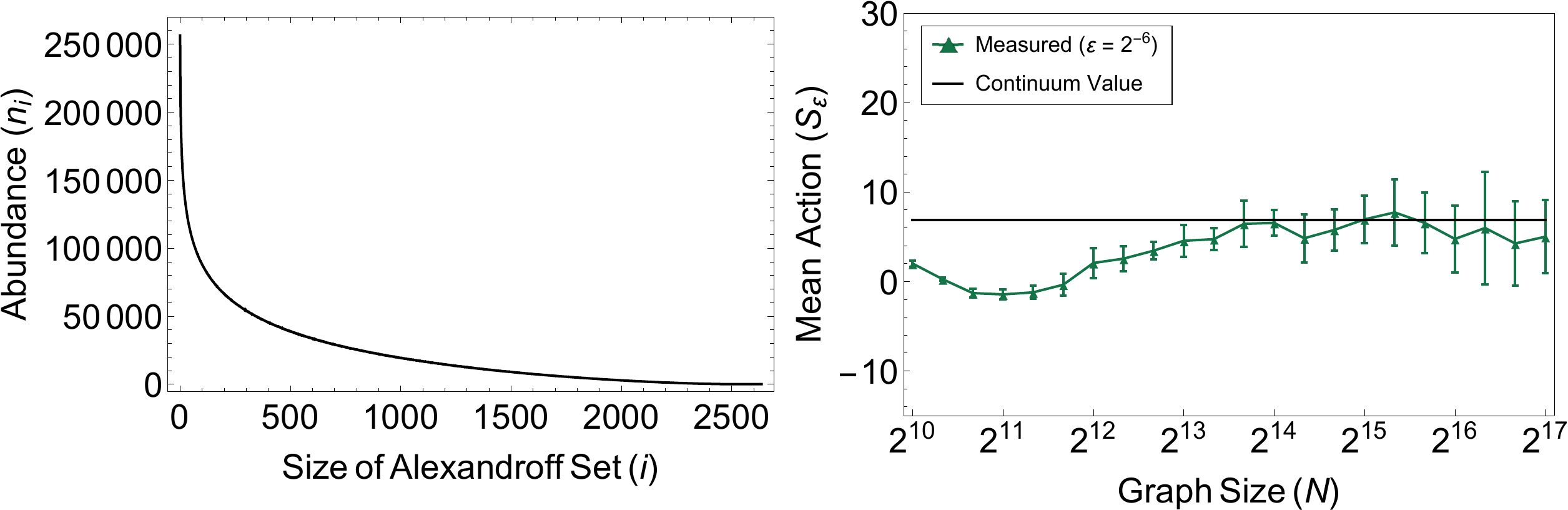}
\caption{{\bf The action in $(1+1)$-dimensional de Sitter spacetime.} The left panel shows the interval abundance distribution for a $(1+1)$-dimensional de Sitter slab with $N=2^{15}$ and $\eta_0=0.5$. The right panel shows the BD action (green) converging toward the EH action (black) as the graph size increases. We take a symmetric temporal cutoff $\eta_0=\pm0.5$ and a small smearing parameter $\varepsilon=2^{-6}\ll 1$ so the onset of convergence appears as early as possible. Remarkably, the terms in the series~\eqref{eq:s_smeared} are several orders of magnitude larger than the continuum result $S\approx 6.865$, yet the standard deviation about the mean is quite small, shown by the error bars in the second panel. The error increases with the graph size because the smearing parameter $\varepsilon$ is fixed while the discreteness scale $l=\sqrt{V/N}$ decreases. All data shown are averaged over ten graphs.}
\label{fig:action}
\end{figure*}

\subsection{MPI Optimization: Load Balancing}
\label{sec:mpi_balanced}
The MPI algorithm described in the previous section grows increasingly inefficient when the pairwise partial vector product operations are not load-balanced across all computers. In Algorithm~\ref{alg:cardinalities}, there is a {\tt continue} statement which can dramatically reduce the runtime when the subgraph studied by one computer is less dense than that on another computer. When the entire adjacency matrix fits on all computers, this is easily addressed by identifying a random graph automorphism by performing a Fisher-Yates shuffle~\blue{\cite{fisher1948}} of labels. This allows each computer to choose unique random pairs, though it introduces a small amount of overhead. \par

On the other hand, if the adjacency matrix must be split among multiple computers, load balancing is much more difficult. If we suppose that in a four-computer setup the {\tt for} loops on two computers finish long before those on the other two, it would make sense for the idle computers to perform possible memory swaps and resume work rather than remain idle. The dynamic design in Figure~\ref{fig:mpi_flowchart} addresses this flaw by permitting transfers to be performed independently until all operations are finished. \par

The primary difficulty with such a design is that for this problem, MPI calls require all computers to listen and respond, even if they do not participate in a particular data transfer. The reason for this is that the temporary storage used for an individual swap is spread across all computers to minimize overhead and balance memory requirements. Therefore, each computer uses two POSIX threads: a master thread listens and responds to all MPI calls, and also monitors whether the computer is active or idle with respect to action calculations, while a slave thread performs all tasks related to those calculations. A shared flag variable indicates the active/idle status on each computer. \par

As opposed to static MPI action algorithm, where whole permutations are fundamental, buffer pairs are fundamental in the load-balanced implementation. This means there is a list of unused pairs as well as a list of pairs available for trading, i.e., those pairs on idle computers. When two computers are both idle, they check to see if a buffer swap would give either an unused pair, and if so they perform a swap. After a swap to an unused pair, the computer moves back from an idle to an active status. 

\section{Simulations and Scaling Evaluations}
\label{sec:simulations}
\subsection{Spacetime Region Considered}
In benchmarking experiments, we choose to study a $(1+1)$-dimensional compact region of de Sitter spacetime. The de Sitter manifold is one of the three maximally symmetric solutions to Einstein's equations, and it is well-studied because its spherical foliation has compact spatial slices (i.e., no contributing boundary terms), constant curvature everywhere, and most importantly, a non-zero value for the action. We study a region bounded by some constant conformal time $\eta_0$ so that the majority of elements, which lay near the minimal and maximal spatial hypersurfaces, are connected to each other in a bipartite-like graph.  \par

The $(1+1)$-dimensional de Sitter spacetime using the spherical foliation is defined by the metric
\begin{equation}
\label{eq:ds_metric}
ds^2 = \sec^2\eta(-d\eta^2 + d\theta^2)\,,
\end{equation}
and volume element $dV = \sec^2\eta\,d\eta\,d\theta$. This foliation of the de Sitter manifold has compact spatial slices, meaning the manifold has no timelike boundaries. Elements are sampled using the probability distributions $\rho(\eta|\eta_0) = \sec^2\eta/\tan\eta_0$ and $\rho(\theta) = 1/2\pi$, so that $\eta\in[-\eta_0,\eta_0]$ and $\theta\in[0,2\pi)$. Finally, the form of~\eqref{eq:ds_metric} indicates elements are timelike-separated when $d\theta^2 < d\eta^2$, i.e., $\pi - |\pi - |\theta_1 - \theta_2|| < |\eta_1 - \eta_2|$ for two particular elements with coordinates $(\eta_1,\theta_1)$ and $(\eta_2,\theta_2)$. This condition is used in the CUDA kernel which constructs the causal matrix in the asynchronous GPU linking algorithm. \par

We expect the precision of the results to improve with the graph size, so we study the convergence over the range $N\in[2^{10},2^{17}]$ in these experiments. Larger graph sizes are typically used to study higher-dimensional spacetimes and, therefore, will not be considered here. We choose a cutoff $\eta_0=0.5$ in particular because for $\eta_0$ too small we begin to see a flat Minkowski manifold, whereas for $\eta_0$ too large, a larger $N$ is needed for \blue{convergence} since the discreteness scale $l=\sqrt{V/N}$ is larger.

\subsection{Convergence and Running Times}
Initial experiments conducted to validate the BD action show that the interval abundance distribution takes the form as in manifold-like causal sets \blue{(versus in Kleitman-Rothschild partial orders)}~\cite{glaser2013}, and that the mean begins to converge to the EH action around $N\gtrsim 2^{14}$, Figure~\ref{fig:action}. The Ricci curvature for the constant-curvature de Sitter manifold is given by $R=d(d+1)$ so that the EH action is simply
\begin{equation}
S_{EH}=\frac{d(d+1)}{2}V(\eta_0) = 4\pi\tan\eta_0\,.
\end{equation}
\blue{We note that the standard deviation $\sigma_S$ in Figure~\ref{fig:action}(right) increases as $O(\sqrt{N}$) because we have chosen to keep the smearing parameter $\varepsilon$ fixed as $N$ increases, which is the more common practice, but if we had instead chosen $\varepsilon\to\varepsilon/\sqrt{N}$, then $\sigma_S$ would go $0$ as $N\to\infty$~\cite{benincasa2013}}. 
While normally one would need to consider the Gibbons-Hawking-York boundary terms which contribute to the total gravitational action, it is known that spacelike boundaries do not contribute to the BD action~\cite{buck2015} and the codimension-2 boundary does not contribute, since the BD action violates the Lorentzian Gauss-Bonnet Theorem~\cite{law1992, benincasa2011b}. \par
\begin{figure*}[pt]
\centering
\includegraphics[width=\linewidth]{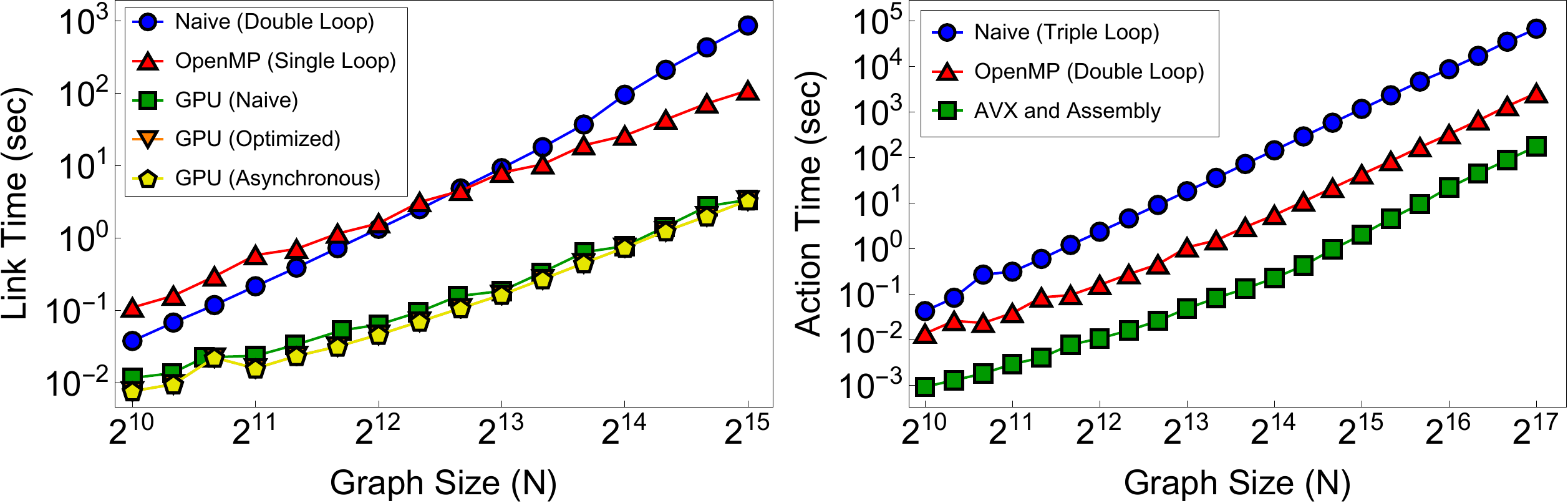}
\caption{{\bf Performance of the linking and action algorithms.} We benchmark the $O(N^2)$ node linking algorithm (left) and the $O(N^3)$ action algorithm (right) over a wide range of graph sizes. The left panel shows moving from a sparse (blue) to a dense (red) representation improves the scaling of the linking algorithm, though it can still take several minutes to generate causal sets of modest size. When the NVIDIA K80m GPU is used, we find a dramatic speedup compared to the original implementation, which allows us to generate much larger causal sets in the same amount of time. We find the three variations of the GPU algorithm (green, orange, yellow) provide nearly identical run times. The right panel shows the benefits of using both OpenMP and AVX instructions to parallelize. The optimal OpenMP scheduling scheme varies according to the problem size, though in general a static schedule is best, since it has the least overhead.}
\label{fig:link_action_times}
\end{figure*}

\begin{figure*}[pb]
\centering
\includegraphics[width=\linewidth]{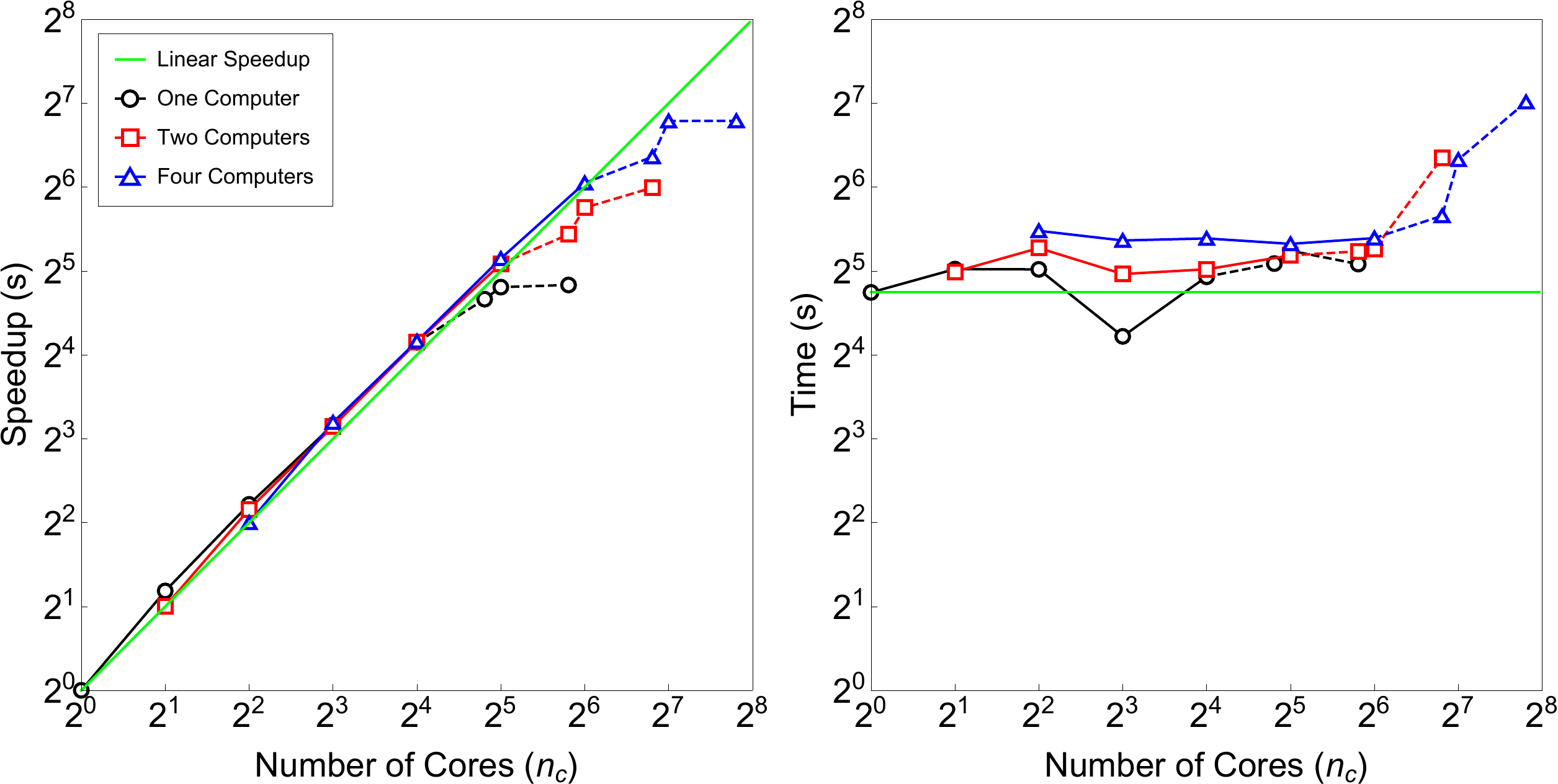}
\caption{{\bf Strong and weak scaling of the action algorithm.} The action algorithm exhibits nearly perfect strong and weak scaling, shown by the straight green lines in each panel. The {\tt for} loop in Algorithm~\ref{alg:cardinalities} is parallelized using OpenMP, while the partial inner product is parallelized using AVX. When multiple computers are used, pairs identified by the loop are evenly distributed among all computers. We find the best speedups when the total number of cores used is a power of two and hyperthreading is disabled (solid lines). When we use all 28 physical cores, or we use 32 or 56 logical cores in our dual Xeon E5-2680v4 CPUs, we find a modest increase in speedup (dashed lines). In the right panel, the runtime should remain constant while the number of processors is increased as long as the amount of work per processor remains fixed. The constant increase in runtime when more computers are added is likely due to a high MPI communication latency over a 10Gb TCP/IP network.}
\label{fig:benchmarking}
\end{figure*}
These calculations are extremely efficient when the GPU is used for element linking and AVX is used on top of OpenMP to find the action (Figure~\ref{fig:link_action_times}). The GPU and AVX optimizations offer nearly a 1000x speedup compared to the naive linking and action algorithms, which in turn allows us to study larger causal sets in the same amount of time. The decreased performance of the naive implementation of the linking algorithm, shown in the first panel of Figure~\ref{fig:link_action_times}, is indicative of the extra overhead required to generate sparse edge lists for both future and past relations. There is a minimal speedup from using asynchronous CUDA calls because the memory transfer time is already much smaller than the kernel execution time. \par

\subsection{Scaling: Amdahl's and Gustafson's Laws}
We analyze how Algorithm~\ref{alg:cardinalities} performs as a function of the number of CPU cores ($n_c$) to show both strong and weak scaling properties (Figure~\ref{fig:benchmarking}). Amdahl's Law, which measures strong scaling, describes speedup as a function of the number of cores at a fixed problem size~\blue{\cite{amdahl1967}}. Since no real problem may be infinitely subdivided, and some finite portion of any algorithm is serial, such as cache transfers, we expect at some finite number of cores the speedup will no longer substantially increase when more cores are added. In particular, strong scaling is important for Monte Carlo experiments, where the action must be calculated many thousands of times for smaller causal sets. We find, remarkably, a superlinear speedup when the number of cores is a power of two and hyperthreading is disabled, shown by the solid lines. The dashed lines in Figure~\ref{fig:benchmarking} indicate the use of 28, 32, and 56 logical cores on dual 14-core processors. \par

We also measure the weak scaling, described by Gustafson's Law~\blue{\cite{gustafson1988}}, which tells how runtime varies when the problem size $N^3$ per processor $P$ is constant (Figure~\ref{fig:benchmarking}(right)). This is widely considered to be a more accurate measure of scaling, since we typically limit our experiments by the runtime and not by the problem size. Weak scaling is most relevant for convergence tests, where the action of extremely large graphs must be studied in a reasonable amount of time. Our results show nearly perfect weak scaling, again deviating when the number of cores is not a power of two or hyperthreading is enabled. We get slightly higher runtimes overall when more computers are used for two reasons: the computers are connected via a 10Gb TCP/IP cable rather than Infiniband and the load imbalance becomes more apparent as more computers are used.  Since the curves have a nearly constant upward shift, we believe the likely explanation is the high MPI latency. For each data point in these experiments, we ``warm up'' the code by running the algorithm three times, and then record the smallest of the next five runtimes. All experiments were conducted using dual Intel Xeon E5-2680v4 processors running at 2.4 GHz on a RedHat 6.3 operating system with 512 GB RAM, and code was compiled with nvcc 8.0.61 and linked with g++/mpiCC 4.8.1 with Level 3 optimizations enabled.

\section{Conclusions}
\label{sec:conclusion}
By using low-level optimization techniques which take advantage of modern CPU and GPU architectures, we have shown it is possible to reduce runtimes for causal set action experiments by a factor of 1000. We used OpenMP to generate the element coordinates in parallel in $O(N)$ time and used the GPU to link elements much faster than with OpenMP. By tiling the adjacency matrix and balancing the amount of work each CUDA thread performs with the physical cache sizes and memory accesses, we allowed the GPU to generate causal sets of size $N\gtrsim 2^{20}$ in just a few hours. We developed the efficient and compact {\tt FastBitset} data structure to overcome limitations imposed by other similar data structures, and implemented ultra-efficient intersection, bit counting, and inner product methods using assembly in Algorithms~\ref{alg:partial_intersection},~\ref{alg:popcnt}, and~\ref{alg:cardinalities}. The MPI algorithms described in Sections~\ref{sec:mpi_static} and~\ref{sec:mpi_balanced} provide a rigorous protocol for asynchronous information exchange in the most efficient way when the adjacency matrix is too large to fit on a single computer. Finally, we demonstrated superlinear scaling of the action algorithm with the number of CPU cores, indicating that the code is well-suited to run in its current form on large computer clusters. 

\section*{Acknowledgments}
We thank J.~Chartrand, D.~Kaeli, C.~Orsini, D.~Rideout, N.~Roy, S.~Surya, and P.~Whitford for useful discussions and suggestions. This work was supported by NSF grants CNS-1442999, CNS-1441828, and IIS-1741355.

\section*{References}
\bibliographystyle{elsarticle-num}
\bibliography{paper}

\end{document}